\documentclass[twocolumn,showpacs,amsmath,amssymb,prl]{revtex4}
\usepackage{graphicx}
\usepackage{bm}
\usepackage{natbib}

\begin{document}

\title{$\lambda$-prophage induction modeled as a cooperative failure mode of lytic repression}

\author{Nicholas Chia$^{1,2}$}

\author{Ido Golding$^{2}$}

\author{Nigel Goldenfeld$^{1,2}$}

\affiliation{$^{1}$Institute for Genomic Biology, University of Illinois
at Urbana-Champaign, 1206 West Gregory Drive, Urbana, IL 61801}

\affiliation{$^{2}$Center for the Physics of Living Cells and Loomis Laboratory of Physics, University of Illinois at
Urbana-Champaign, 1110 West Green Street, Urbana, IL 61801}

\date{\today}

\begin{abstract}
We analyze a system-level model for lytic repression of
$\lambda$-phage in \textit{E. coli} using reliability theory, showing
that the repressor circuit comprises 4 redundant components whose
failure mode is prophage induction. Our model reflects the specific
biochemical mechanisms involved in regulation, including long-range
cooperative binding, and its detailed predictions for prophage
induction in \textit{E. coli} under ultra-violet radiation are in
good agreement with experimental data.
\end{abstract}



\maketitle

Viruses were one of the first biological systems to attract the
attention of physicists\cite{ellis1939gb}. Despite their apparent
simplicity, it has become increasingly
clear\cite{casjens2003pab,goldenfeld2007cbs} that they provide a
window into the full systems complexity of the
cell\cite{ptashne2004gsp}, as well as representing a major
evolutionary\cite{brussow2004pae} and ecological
force\cite{weinbauer2004vdm} affecting all three domains of life.
The viruses that infect bacteria are known as bacteriophages (or
phages). When some phages infect a bacterial cell, there can be two
possible outcomes or pathways\cite{ptashne2004gsp}.  In the lytic
pathway, the phage hijacks the cell's machinery to replicate itself
many times, and to release the replicates by breaking open or lysing
the cell.  In the lysogenic pathway, the phage integrates its genome
into that of the host microbe, becoming a prophage, but otherwise
does not damage the cell. This lysogenic state is very
stable\cite{baek2003sca}; however, an insult to the cell through,
for example, starvation or exposure to ultra-violet (UV)
radiation\cite{luria1947uib} can trigger a process known as prophage
induction\cite{oppenheim2005}: the prophage is excised from the
cell's genome, and viral replication occurs leading to cell lysis.
The most well-studied lysogenic system is the bacteriophage
$\lambda$, or $\lambda$-phage, which infects \textit{Escherichia
coli}. Understanding the lysis-lysogeny system in detail is
important, because this system is one of the simplest examples of a
gene regulatory network\cite{hasty2001csg}---a pervasive and
fundamental form of biological organization and function, whose
principles are still being elucidated.  Although there has been
considerable interest recently in the role of
stochasticity\cite{arkin1998ska} in the switching behavior between
lytic and lysogenic states as part of the phage
life-cycle\cite{ackers1982qmg,shea1985ocs,aurell2002spp,lipshtat2006gts,bialek2008csa},
here we focus on UV prophage induction, where a different mechanism
is involved.

UV prophage induction experiments exhibit threshold
behavior\cite{Kneser1966}, in which the fraction of induced lysogens,
(i.e., prophage containing cells) rapidly increases as a function of UV
dose. Under typical laboratory growth conditions, the fraction of
induced lysogens versus the UV dosage obeys a power law with a power
very close to 4\cite{Kneser1966}. Power law behaviors of this type can
arise in several ways: (i) as an event caused by 4 independent hits on
a \lq\lq target" (target theory\cite{stent1963mbb}); or (ii) a chemical
equilibrium reaction involving a substrate bound to 4 chemical species,
and quantified by the empirical Hill equation for chemical
kinetics\cite{hill1910pea,rosenfeld2005grs}. Target theory and chemical
kinetics could provide a way of understanding how UV dose curves can
yield an exponent of 4, but have little connection to the biochemical
regulatory mechanisms of $\lambda$-phage lytic
repression\cite{oppenheim2005,court2007nlb}. An alternative perspective
is to view UV induction in the framework of the standard stochastic
model of lytic repression\cite{shea1985ocs,aurell2002spp} with adjusted
rate constants. While these models are informed by the biochemistry,
the mapping to UV prophage induction remains dependent on unknown
parameters.

In this Letter, we show how the emerging understanding of the role of
DNA loops and long-range cooperative
binding\cite{dodd2004clr,dodd2007mti} in the biochemical picture of
lytic repression can account quantitatively for the phenomenology of
prophage induction. Our approach is to abstract the biochemistry into a
systems-level description, in which the lytic repressor circuit is
represented as a device comprised of a number of redundant elements and
one failure mode, lysis. This allows us to draw connections between the
biochemical regulatory mechanism and reliability
theory\cite{gertsbakh1984,gavrilov2001rta,gavrilov2006a} and also
predict the characteristic power law for UV prophage induction.


\begin{figure}[t]
\begin{center}
\includegraphics[width=0.9\columnwidth]{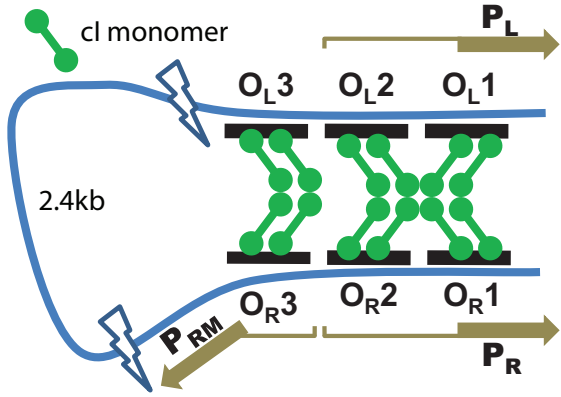}
\caption{\label{switch1} Schematic diagram of the $\lambda$-phage lytic
repression system.  Attachment of RNA polymerase to promoter regions,
indicated here by arrows, lead to gene expression. $P_R$ and $P_L$ lead to
expression of genes in the lytic pathway. However,
CI dimer binding at $O_R1$ or $O_R2$ blocks transcription of $P_R$ while CI binding
at $O_L1$ and $O_L2$ block transcription of $P_L$.
$O_R3$ likewise regulates the transcription of genes by the promoter region
$P_{RM}$ while CI bound to $O_R2$ promotes transcription of genes from $P_{RM}$.
Dimers of CI are capable of forming stable quadramers when
attached to adjacent sites such as at $O_R1$ and
$O_R2$\cite{ackers1982qmg,shea1985ocs} (modeled by \cite{vilar2006mdl}).
Furthermore, $O_R1$-$O_R2$
and $O_L1$-$O_L2$ quadramers can form a stable octamer in a long
range interaction typically spanning $~2.4$ kb, but up to 3.8
kb\cite{revet1999fdl,dodd2001olc}.
}
\end{center}
\end{figure}

\begin{figure}[t]
\begin{center}
\includegraphics[width=0.9\columnwidth]{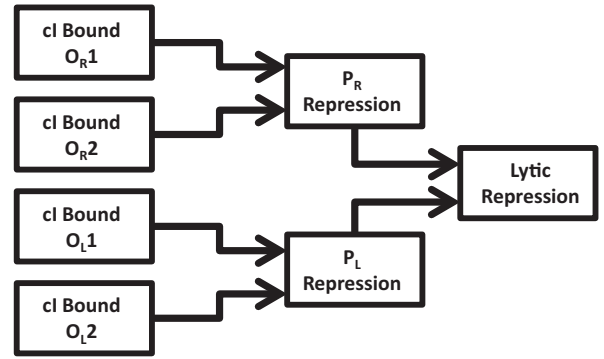}
\caption{\label{switch2} $\lambda$-phage lytic repression regulatory
circuit. CI bound to $O_R1$ or $O_R2$ blocks expression of lytic genes under
the control of $P_R$. CI bound to $O_L1$ or $O_L2$ blocks expression
of lytic genes under the control of $P_L$. Since both sets of genes under the
control of $P_R$ and $P_L$ are required, blocking expression of either
effectively represses lysis. Thus, each of the 4 CI bound components
blocks lysis when intact. These can be seen as redundant elements that perform
the same task.}
\end{center}
\end{figure}

\textit{Biochemistry of lytic repression:-} Fig.~\ref{switch1} illustrates
a widely accepted model of the lytic repressor switch in
$\lambda$-phage (for a review see \cite{ptashne2004gsp}).
The lytic repressor molecule CI dimerises and binds to specific DNA
sites in the $O_L$ and $O_R$ control boxes, $O_L1$ and $O_L2$, blocking
expression of genes under the control of the $P_L$ promoter. In the
$O_R$ control box, $O_R1$ and $O_R2$ regulate $P_R$. Since only free
$O_R1$ and $O_R2$ sites allow the expression of $P_R$ controlled genes,
CI binding at either the $O_R1$ or $O_R2$ suffice for repression of
$P_R$. Fig.~\ref{switch2} sketches the relationship between these 2
$O_R$ sites and $P_R$ expression. The same applies to the role of
$O_L1$ and $O_L2$ in suppressing $P_L$. Also, as drawn in
Fig.~\ref{switch2}, expression of genes under the control of both $P_L$
{\em and} $P_R$ promoters lead to lytic development of the
prophage\cite{dodd2005rgr}. Derepression of all of the 4 binding sites
results in lysis while bound CI dimers at any site block the lytic
pathway. In UV induction, RecA-mediated autocleavage of CI monomers
deprives the binding sites of available CI dimers. RecA-mediated
autocleavage can happen once RecA is activated as part of the host SOS
response to DNA damage\cite{little1982srs}.


\textit{Abstraction of the lytic repression circuit:-} Prophage
induction can be understood as the failure of the lytic repression
circuit, which consists of 4 redundant components that each prevent
lysis.  Each has a failure rate $\mu_i$ ($i=1\dots 4$) per UV dose
$x$ and a corresponding survival probability $p_i = \exp
(-\mu_i x)$.
Fig.~\ref{switch2} diagrams the relevant aspects of the $\lambda$-phage
lytic repressor regulatory system. Each component consists of a lytic
repressor CI dimer bound to one of 4 specific DNA sites, i.e., either
$O_R1$, $O_R2$, $O_L1$, or $O_L2$. Each site regulates the expression
of genes essential to the lytic pathway by its influence on either the
promoter $P_R$, by $O_R1$ and $O_R2$, or $P_L$, by $O_L1$ and $O_L2$.
Thus, these 4 components have redundant functionality, i.e., repressing
lysis. Since suppression of genes under the control of either promoter
keeps lysis in check, only damage to the final component results in
lysis.

{\it Reliability theory of the repressor circuit:-} To understand the
failure rate of the system, i.e., the fraction of cells lysed, note
that the probability of failure, for UV dose $x$, is
$1-p_i$ for each of the 4 redundant components (the 4 CI dimers bound
to $O_R1$, $O_R2$, $O_L1$, and $O_L2$) in the lytic repression system
(see Fig.~\ref{switch2}).
In general, the probability of failure as a function of total UV dose,
$F$, for a system of $n$ redundant components is
\vspace{-6mm}
\begin{center}
\begin{equation}
F = \prod_{i=1}^n(1-p_i)\label{eq:Fp}.
\end{equation}
\end{center}
We model the effects of radiation on the failure rate of the
lytic repression system, by assuming that near the
threshold $x_c$, $p_i = \exp[-\mu_i (x-x_c) + O((x-x_c)^2)]$.

By taking measurements of the fraction of failed systems and
Eq.~\ref{eq:Fp}, the number of redundant elements in the system can
be deduced. The fraction of failed units is then given by
\vspace{-6mm}
\begin{center}
\begin{equation}
F(x) = \prod_{i=1}^n(1-\exp[-\mu_i x]) \approx (\mu x)^n
\quad \quad (\mu x \ll 1)
\label{eq:FT}
\end{equation}
\end{center}
where $\mu = (\prod_{i=1}^n \mu_i)^{1/n}$ is an effective failure rate.
The UV prophage induction curve describes the fraction of cells lysed
as a function of UV dose $x$ and is predicted to follow Eq.~\ref{eq:FT}
with $n=4$. In other words, the fraction of cells lysed can be computed
from the effective failure rate $\mu$ for the 4 CI dimer bindings at
$O_R1$, $O_R2$, $O_L1$, and $O_L2$. The effective rate of failure $\mu$
varies between different experimental systems and depends on a
number of parameters. For example, different {\it E. coli} hosts may
exhibit varying levels of RecA
activity\cite{dutreix1985eip,little1982srs}. Alternatively, mutant {\it
cI} alleles may offer operator site binding
affinities\cite{nelson1985lrm}. Below, we test Eq.~\ref{eq:FT} against
data from experiments on radiative induced lysis, and extract $\mu$.

\begin{figure}[t]
\begin{center}
\includegraphics[width=\columnwidth]{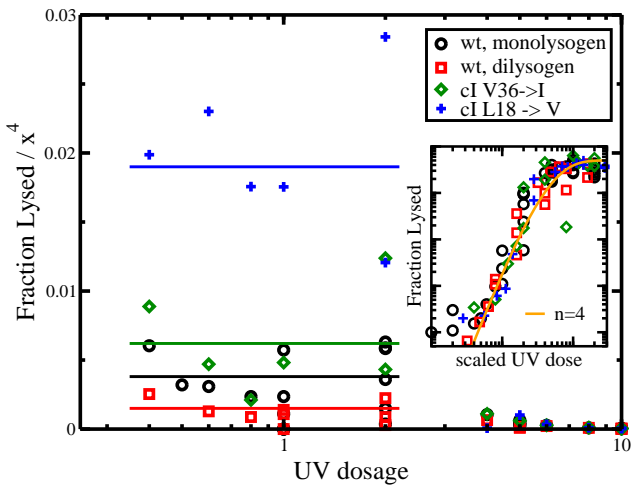}
\caption{\label{data}The UV prophage induction curve.
$F/(x^4)$ versus UV dosage $x$. 
The data is taken from different $\lambda$ strains. Wildtype prophages were
integrated as monolysogens and dilysogens with one or two inserted $\lambda$
genomes, respectively. All 4 induction curves scale approximately as a
power of 4 for small UV doses ($0.2$-$2$ J/m$^2$). (Inset) Log-log plot
of the same data. Curves are right- and left-shifted so that they overlay
each other. The line represents Eq.~\ref{eq:Fp} for $n=4$.}
\end{center}
\end{figure}

{\it Experimental tests:-} In order to measure the fraction of lysogens
induced as a function of UV dose, we followed standard
protocols\cite{little1999rgr}. Briefly, exponentially growing lysogenic
cells were harvested and resuspended in buffer. The cells were then
irradiated by a germicidal UV lamp in dim ambient light for a range of
doses at $\sim 1$ J/m$^2$/s. After irradiation, aliquots were diluted into
growth medium, shaken for 2 h at 37°C in the dark, treated with CHCl$_3$
and titered for plaque forming units.

In our abstraction of the lytic repression system, we noted four redundant
components, as shown in Fig.~\ref{switch2}. Since this implies $n=4$,
we can test the applicability of Eq.~\ref{eq:FT} to our experimental
results by plotting $F/(x^4)$ against UV dose $x$, as shown in
Fig.~\ref{data}.

In Fig.~\ref{data}, the effective rate of failure $\mu$ manifests
itself as a horizontal line. As shown in Fig.~\ref{data},
experimental data agrees with Eq.~\ref{eq:FT} for a certain range of
UV dosage. Disagreement between theory and experiment occurs at both
low and high UV dosage. The breakdown in both these regimes is
readily interpreted. For extremely low dosage (near zero),
spontaneous lysis events not induced by UV irradiation become the
strongest contributing factor to the failure rate. These failures
come from other events, such as spontaneous RecA
activity\cite{oppenheim2005} and mutations to
$\lambda$-phage\cite{baek2003sca}. At high radiation doses (not
plotted, see \cite{Kneser1966}), the fraction of lysis does not
saturate at one and instead begins falling with respect to UV dose.
Here, damage to the lytic pathway likely results in the inability of
a cell to lyse, either because key components of the lytic pathway
or host metabolism have been crucially damaged. In other words, this
drop is not the effect of a lower failure rate of lytic repression,
but a reflection of the high failure rate of other cellular systems
upon which lysis relies. In these cases, failure of lytic repression
cannot be detected by cell lysis since the lytic response has been
disabled.

{\it Discussion:-} Our model postulates that induction, or a failure
event, is dominated primarily by CI dimer dissociation. Though many
factors, such as DNA damage and RecA activity~\cite{little2005teg},
contribute in principle to the failure of lytic repression, near the
threshold of prophage induction, only the most rapidly-varying
parameter is important. The rapidly-varying CI-dimer operator site
bindings about the threshold point for lytic induction $x_c$
justifies the approximation made in Eq.~\ref{eq:FT} allowing us to
observe the predicted $n=4$ power law.

In contrast, chemical kinetic models assume that since the timescale
of CI dimer dissociation ($\sim 30$ sec\cite{nelson1985lrm}) falls
an order of magnitude below the time required for lysis ($\sim 30$
min), the CI bindings can be regarded as being adiabatically slaved
to CI monomer concentration\cite{shea1985ocs,aurell2002spp}, and
thus not a determining factor in the switch. The prophage induction
curve and its power law behavior then arise from the behavior of the
CI monomer depletion. However, as can be seen from the data
presented in Fig.~\ref{data}, the power law dependence $n=4$ is
robust, and not sensitive to the different strains or experimental
conditions. In our model, the power $n$ reflects the number of
redundant elements---not the properties of the individual
components---and so is robust. This can be directly tested by
manipulating the number of redundant operator sites and measuring
the resultant power law dependence. Ref.~\cite{rosenfeld2005grs}
measured $P_R$ expression as a function of CI concentration. As
shown in Fig.~\ref{switch1}, $P_R$ expression is regulated by 2
bound operator sites, so we predict that $P_R$ expression should be
described by a kinetic curve with an $n=2$ power law dependence,
as has been previously noted\cite{bintu2005trn}.
The kinetic data of ref.~\cite{rosenfeld2005grs} are indeed consistent
with the
prediction we have made here, based on our system-level abstraction
of the underlying biology. This finding supports our view that the
threshold behavior of prophage induction is determined by CI, and
not by other steps in the repressor circuit. To establish this more
conclusively, it would be necessary to check that $P_L$ expression
also exhibits the predicted $n=2$ power law behavior.

Our model implies that the repressor sites are the key to stabilizing
the lysogenic state while the presence of Cro does not play a role in
the switching\cite{svenningsen2005rclambda,atsumi2006rlr} except to
enforce commitment to the developmental
transition\cite{schubert2007csr}. Once the switch has been activated,
it cannot be reversed, due to the role of Cro.  Our model also suggests
a mechanism for abortive induction events that are sometimes
observed\cite{oppenheim2005}. In our model these arise when unblocked
$P_L$ transcribes the genes required for excision (see
\cite{ptashne2004gsp}) while $P_R$ remains blocked.

As shown by Fig.~\ref{data}, different conditions lead to different
values for $\mu$. We can use reliability theory to anticipate the
trends in variation of $\mu$ between 2 similar experiments. The rate
of component failure depends on a number of variables including RecA
activity, CI concentration, binding strength of repressor sites, and
stability of CI to autocleavage. Here, damage to redundant
components corresponds to the dissociation of CI dimers from
$O_L1/2$ and $O_R1/2$ sites (see Fig.~\ref{switch2}).

\begin{table}
\caption{\label{table}List of predicted and observed changes in
inducibility tabulated according to mutation, either in $\lambda$ or the {\it E. coli}
host. $\uparrow$ indicates that the change results an increase in $\mu$, while $\downarrow$
indicates a decrease. Predictions based on reliability theory match with currently available
data.}
\begin{ruledtabular}
\begin{tabular}{llccl}
Strain & Phenotype & Theory & Data & Ref. \\ \hline 
$\lambda$ cI $ind^{s}$-1 & faster CI cleavage & $\uparrow$ & $\uparrow$ & \cite{dutreix1985eip} \\
{\it lexA51 recA441} & increased RecA activity & $\uparrow$ & $\uparrow$ & \cite{dutreix1985eip} \\
$\lambda$ cI $ind$543 & stronger CI dimerization & $\downarrow$ & $\downarrow$ & \cite{dutreix1985eip} \\
$\lambda$ cI Y210$\rightarrow$N & disrupts CI dimer-dimer & $\downarrow$ & $\downarrow$ & \cite{babic2007cdb}\\[-1mm]
 & cooperative binding & & \\
$O_R2^*$ & weaker $O_R2$-CI binding &  $\uparrow$ & $\uparrow$ & \cite{rosenfeld2005grs}\footnotemark[1]\\
\end{tabular}
\end{ruledtabular}
\footnotemark[1]{Measured $P_R$ expression}
\vspace{-2mm}
\end{table}

Table~\ref{table} lists theoretical predictions for the variation in
the failure rate of the lytic repressor or inducability arising from
possible laboratory manipulations of the rate of failure $\mu$. As
expected, increasing RecA or rate of CI autocleavage leads to
increased failure rates while an increased number of CI dimers slows
down the rate of failure. Increasing operator site binding strengths
through cooperative binding or operator site mutations lowers the
rate of failure. In summary, weakening the CI operator site bindings
results in an increased failure rate for the lytic repressor, while
increasing the probability of those binding events will have the
opposite effect.

We thank David Reynolds and Carl Woese for helpful discussions. IG
wishes to express his deep gratitude to Edward Cox, at whose lab the
phage induction experiments began. IG is partially supported by NIH
grant R01-GM082837-01A1. NC was partially supported by Department of
Energy grant No. DOE-2005-05818, and the IGB Postdoctoral Fellows
Program. This material is based upon work supported in part by
the National Science Foundation under Grant No. 082265, PFC: Center
for the Physics of Living Cells.

\bibliography{bibfileL}

\end{document}